\documentclass{article}
\begin{document}

STATISTICAL PHYSICS OF  DARK AND NORMAL MATTER DISTRIBUTION  IN GALAXY FORMATION :
\qquad DARK MATTER LUMPS AND BLACK HOLES IN CORE AND HALO OF GALAXIES.

\qquad \qquad\qquad\qquad \qquad  Ajay Patwardhan

\qquad \qquad Physics Department, St Xaviers college, Mumbai

\qquad \qquad Visitor, Institute of Mathematical Sciences, Chennai

\qquad \qquad\qquad\qquad \qquad ABSTRACT

In unified field theory the cosmological model of the universe has
supersymmetric fields. Supersymmetric particles as dark and normal  matter in
galaxy clusters have a phase separation. 

Dark matter in  halos have a statistical physics equation of
state. Neutralino particle gas with gravitation can have a collapse of
dark matter lumps. 

A condensate
phase due to boson creation by annhillation and exchange can occur at
high densities.
The collapse of the boson condensate, including neutralinos, into the Schwarzschild radius creates
dark matter black holes. 

Microscopic dark matter black holes can
evaporate with Hawking effect giving gamma ray bursts and create a
spectrum of normal particles. 

The phase separation of normal and dark
matter in galaxy clusters and inside galaxies is given by
statistical physics.  

\section
\qquad\qquad\qquad \qquad \qquad INTRODUCTION

The current undertanding of cosmology has dark energy, dark matter and
normal matter including electromagnetic radiation as contents of the
universe; all arising from an inflationary phase. 

A unified field
theory [Refs 9] can have scalar fields coupled to others and create
spacetime geometry and all forms of particles. 

The dark energy gets associated with vaccuum and Casimir energy. It is
associated with the inflationary scalar field and with the
cosmological constant of spacetime geometry in various epochs of the
universe. 

It can also be a fluid such as Chaplygin gas or phantom or
tachyonic particles. Cosmological observational evidence gives a' best
estimate'equation of state and models give some properties.

Dark energy has a phase separation from the dark and normal matter around 10
million years from big bang; with voids full of repulsive dark energy
and collapsing 3,2,1 dimensional structures of normal and dark matter that appear as filaments
with point like galactic nodes at the scale of gegaparsecs down to 100
megaparsecs.

The inhomogeneity structure formation from WMAP time of $< 1$
million years to large scale structure formation of 100s of million
years includes this epoch.

The dark matter and normal matter particles are interspersed in the
pregalactic formation epoch. In the process of formation of galaxies,
the interspersed voids contain dark matter excess over normal matter.

Supersymmetric particles of dark matter decay and the neutralino is a
lowest mass 
remnant particle. A relativistic gas at low temperatures is formed as
can be understood as follows.

The SUSY particles do not thermalise with the
background electromagnetic radiation and they do not interact with
photons once the electric, colour and flavour charge free states are
produced. They do not interact with the leptonic and baryonic normal
matter and hence do not loose energy by collisions. 

They remain
relativistic at the energy when the last decay to neutralino
occurred. As neutralinos have very small interaction crosssections
they will have negligible thermal energy or temperature. 

The dark
matter is interspersed and phase separated from normal matter. And
there is excess dark matter over normal matter of 3 to 8 times.

Normal matter goes through the well known stages of baryon, nuclei, atom
,molecule, dust formation which interacts with electromagnetic
radiation of appropriate frequencies in each epoch of the expanding
universe.

 This matter also interacts among itself and a thermal
distribution is attained. This leads to a non relativistic gas of
normal matter  at low temperatures.

The neutralino gas with particle rest energy 300 to 1400
GeV and the baryonic/ leptonic gas with 1GeV to 1 MeV rest energy have
two different intrinsic temperatures or average thermal energies and
they phase separate.

 The core of galaxy is formed with a AGN or
supermassive black hole and a spherical  halo is formed with dark matter; with
the normal matter residing in the potential well in between. 

This
becomes visible as the luminous part of the galaxy in some
wavelengths. Star formation and other processes eventually give
radiation at all wavelengths from the galaxy.

Through out this process that lasts from 100 million years to > 10
billion years from origin; the dark matter neutralino gas is a
relativistic fermionic gas going from ultrarelativistic to non relativistic, in a self created  gravitational
potential.

It can create an effective gravitational potential in the galaxy as well
as local fluctuations that grow in density and collapse to form dark
matter lumps with a wide range of mass spectrum. 

Some of these become dark matter black holes; as at short range and
 high densities the interaction crosssection of neutralinos increases 
 due to exchange bosons including Higgs boson. 

The Bose condensation
 of Higgs and exchange W,Z bosons along with the gravitational collapse of the
 fermionic  neutralino lumps can take all the dark matter  mass into
 the Schwarzschild radius, creating a dark matter black hole.

A mass spectrum of such dark matter black holes gives the possibility
that microscopic black holes would be able to quantum evaporate by
Hawking effect. This may lead to a flux of gamma rays and normal
matter  particles. Such events have been reported from galaxy halos,
but the data is still insufficient and not conclusive.

 The entire
process depends on the statistical physics of the dark and normal
matter gas including gravitational effects at long range and particle
interactions at short range. For black hole formation general
relativistic effects have to be included.

In an earlier work [Ref 9 ] the formation and  quantum evaporation of
primordial black holes was considered.

In this paper the statistical
physics of dark matter is worked out and the various conditions of
galaxy formation and galaxy halos with dark matter black holes is
obtained.

The observational tests for the interspersed dark matter and
the evaporating dark matter black holes are mentioned. This work should
complement the growing tests for dark matter in galaxy cores, in
galaxy halos and galaxy clusters.

As a comparison with other work on this problem consider the results
of the papers [Refs 1 to 6 ] (a) density functions in halos ex. [1]     (b)equations
of state ex. [2] (c)  annhillation crossections ex. [3]  topics.

\section
\qquad STATISTICAL PHYSICS OF NORMAL AND DARK MATTER IN GALAXIES:

The statistical physics of Fermionic and
Bosonic particles is developed for normal and dark matter. For the
dark matter a neutraliono $ \chi^{0}_{1}$ particle is taken as the
remnant of SUSY decays.

The relativistic case of fermionic neutralino gas with limits of non
relativistic and ultrarelativistic gas, the low and high temperature
limits and low and high density limits is considered in the presence of self
gravitation effects and interactions  with exchange bosonic particles.

We begin with the standard expressions for free non interacting
identical particles.

$< N> = \Sigma \frac{g(E)}{(z^{-1} exp (\beta E) + a) }$

$ <E> = \Sigma  \frac{g(E)E}{(z^{-1} exp(\beta E ) + a )}$

$< \frac{P V } { K T}> = \Sigma a^{-1} g(E) Ln ( 1 + a z exp(-\beta E)
)$

where $ a $ is $ +1$ for Fermions, and $ -1$ for Bosons,

 $ z $ is
fugacity.$ z = exp(\mu/KT)$

The supersymmetric matter of Fermionic gas of very weakly interacting neutralinos exists in a self created 
gravitational potential at low densities. 

Neutralinos interact with exchange gauge and
Higgs bosons at short range when gravitational collapse causes high
density. 

The gravitational potential that was for low densities  now becomes the effective
potential in the Schwarzschild geometry at high densities.

 The
statistical physics during 
collapse can be described in a comoving coordinate system in this geometry. A 
Bose condensation can be expected for the bosons. This accelerates the
collapse. 

If the entire mass of 
fermions and bosons resides inside the horizon then a dark matter
black hole is formed. Otherwise a dark matter lump  occurs.

Next the
mixture of normal matter and dark matter distribution is
considered. At length scales of 100 megaparsecs this is nearly uniform
with a $ 1:5 $ ratio. 

The phase separation into galaxy clusters and
intergalactic space happens from 10 million years onwards within the
filaments of normal and dark matter formed with voids filled with dark
energy.

 The intergalactic spaces have a low density dark matter and
negligible normal matter. With galaxy formation around 100 million
years onwards the phase separation creates galaxy halos with dark
matter  and interiors with normal matter.

 The question of interspersed
dark matter within the galaxy and the ratio of dark and normal matter
in the galaxy cores that have collapsed into active galactic nuclei as
supermassive black holes is an open question. 

The statistical physics
of the model developed in this paper gives some insight into the phase
separations.

With $ E = mc^{2}[( 1 +( \frac{p}{mc})^{2})^{1/2} - 1]$,

and $( p/mc) = sinh(x) $,

 $ E = mc^{2}( cosh(x) -1) $ ,

 $ \frac{dE}{dp} = c tanh(x)$

$g(E)dE = g(p)dp $ , $ g(p) dp = \frac{ 4 \pi V}{ h^{3} }p^{2} dp $

substituted in equations above gives the relativistic quantities.

  Then pressure is $ P = \frac{ 4\pi m^{4}c^{5}}{h^{3}}
 \int_{0}^{x_{f}} sinh^{4}(x) dx  $

And total internal energy is $ U = \frac{4 \pi V m^{4} c^{5}}{h^{3}}
\int_{0}^{x_{f}}   ( cosh(x) - 1) sinh ^{2} (x) cosh(x) dx $

Consider the relevant approximations:

 Ultrarelativistic normal and dark matter , bosonic and fermionic
with

  $ T = 10 ^{ 12} $ eV and below.

 Then $z = exp(0) = 1$ and

 $
exp(\beta E)= 1 $ .

 The $ z^{-1} exp(\beta E ) = 1 $.

$q(V,T) = \frac{pV}{KT} = \frac{ 8\pi V g_{4} (
  z=1)}{(hc)^{3}\beta^{3}}$

 with $g_{4} (1) = \zeta(4) = \pi ^{4} /90 $

 and $ P = 1/3 U/V $ for the
bosonic particles.

For the fermionic ones it is   
 
$q(V,T) = \frac{pV}{KT} = \frac{ 56\pi^{5} V }{360(hc)^{3}\beta^{3}}$

and higher order terms in $ (\mu/KT)^{2} $

The expressions for number density $ N/V $ are also obtained as 

$ \frac{ 4\pi^{3} V }{3(hc)^{3}\beta^{3}} \mu/KT$

 for Fermionic and 

$ \frac{ 8\pi V \zeta(3) }{(hc)^{3}\beta^{3}}$ for Bosonic case.

 This
is the early epoch of uniformly interspersed dark and normal matter.

The normal matter particles have masses $ 1 MeV$ and $ 1 GeV$ for
leptons and baryons.

 As the temperatures  go below TeV, they
go from being ultrarelativistic to relativistic to non relativistic
 after  those of SUSY matter whose lightest particles of dark matter ,
 neutralinos are mass $ 10 ^{3}  GeV $. 

Hence in the epoch of 100
 million years we could take the dark matter as non relativistic and
 the normal matter as relativistic. However the normal matter couples
 to radiation and equilibrates thermally.

 In the expanding universe as
 the radiation spectrum shifts to longer wavelengths the normal matter
 also becomes non relativistic . In the in between phase due to
 different approximate equations of state and number densities, energy
 densities  arising , there is a phase separation. 

The dark matter
 interacts negligibly with normal matter and not at all with radiation
 . It has few energy loss mechanisms. 

If the temperature becomes much less than eV compared to the rest and
kinetic energies of particles then we could take it to
be zero. 

 Then $z^{-1}exp(\beta E) = 0 $ for the fermionic gas.

 Then
the relativistic gas has

 $ P = \frac{ \pi m^{4} c ^{5}
  A(sinh(p/mc))}{6h^{3} } $

 and $ U =  \frac{ \pi V m^{4} c ^{5}
  B(sinh(p/mc))}{6h^{3} }$

 where $ A $ and $ B$ are Airy functions.

 Two useful limits are for $
 sinh(p/mc) << 1 $ and $ >>1$ 

respectively non relativistic and
 ultrarelativistic  case.

 For the Pressure and total internal energy  given above 
The Airy
functions

 in these limits give :

$ P =  \frac{4\pi p^{5}}{ 15 m h ^{3} } = 2/3 U/V $

 and  $ P = \frac{ \pi c p^{4} } { 3 h^{3} } = 1/3 U/V .$

 For dark matter the mass  m is 1000 times more than for normal matter
 leading to much less pressure and energy density. Also the Fermi
 momentum at which these are evaluated is

 $ p = (\frac{ 3 h ^{3} N}{ 4
   \pi V })^{1/3}.$

This is lower for dark matter because the $ 1: 5 $
 ratio by total mass of normal and dark matter , and $ 1: 1000 $ ratio of
 particle masses implies a $ 200: 1 $ ratio of numbers of normal and
 dark particles.

 Hence a smaller $ N/V $ for dark matter and a lower
 $p$ Fermi momentum too. Hence we expect with $ 200^{1/3} = 6 $
 and a factor $ 10 ^ {3} 6^{5}$ less pressure for dark matter than normal
 matter.

It may be expected then that any accumulation of  a mixture of normal
and dark matter can have a thermodynamic non equilibrium in which the
dark matter would accumulate more rapidly due to gravitational effects
as it has less thermodynamic  pressure
and collapses the widely dispersed mixture into pregalactic lumps and
intergalactic spaces.

 Further the dark matter would collapse to the
core of the pregalactic lump leading to a largely dark matter
supermassive black hole.

 The phase separation of dark matter halos and
normal matter in the potential well between core and halo is one
important model of galaxies. 

However the model of dark matter
interspersed in the galaxy with normal matter is a possibility too. 
If the neutralinos do not thermalise by scattering or radiative
methods then they can continue to move at relativistic speeds as they
were when they were formed. Virialisation is not a definite or
significant process for them. 

Hence they will reside for negligible time
in the filled spaces by normal matter , such as planets and  stars. But
they will fill mostly the much more voluminous void spaces in between
in the 
solar system, star clusters as well as in the galaxy. Using
crosssections for interaction with baryonic matter while passing
throuigh filled regions,  a flux limit for
detection can be setat Earth.

Dark matter is
therefore a relativistic Fermi gas at very low temperatures at the
time of pregalactic lump formation, which
becomes non relativistic over the life time of a galaxy residing
mostly in the halo.

For the collapsing galactic material , at any time there is a volume
in the equations above that decreases in time. The gravitational
potential energy is added to the thermodynamic energy as well as the
gravitational pressure to thermodynamic pressure. Considering a
spherically symmetric model for the halos and lumps in halos,
 the gravitational
potential energy inside the halo sphere is: 

$ \phi (r) = a - b R^{2} $ 

For flat velocity rotation curves a constant potential model is
taken. The central cusp model gives various density functions. Keeping
density as a $ 1/R^{2} $ function gives a mass function  linear in R.

The gravitational pressure is : $ P_{G} =  \frac{G M ^{2}}{4 \pi R^{2} }$

The equilibrium condition then becomes :

$ U + \phi $ minimised and $ P + P_{G} < 0 $ . 

Actually this condition
is dynamical and  leads to collapse as the neutralino gas component
phase separates . 

In the halo density fluctuations around average 
$\rho (R) $ create gravitational accretion . 

Using limits of low number
density and low Fermi momentum for  the
Airy functions, the relativistic low temperature neutralino $ 1000 G e
V $ rest mass gas can
form a mass spectrum of dark matter lumps

 $ R = const M^{-1/3}(1000
GeV ) ^ {-5/3} $.

 These can range from 100 stellar masses to 1/10
stellar masses. If the other limit of high density and large fermi
momentum is taken then

 $ R = const M ^{1/3}(1000 GeV)^{-1/3}$.

\section \qquad COLLAPSE OF DARK MATTER LUMPS IN GALAXY HALOS AND CORE: FORMATION OF BLACK HOLES

 As the
collapse proceeds these dark matter lumps go from stellar radii to
pebble size. The neutralino annhillation cross section increases and
bosons , real and virtual are produced.

The neutralino $ \chi^{0}_{1} $ particles have very low cross section for
interactions, including anhillations, that produce normal particles --
particularly Z, W and H bosons. This effect of weakly interacting massive
particles ( WIMPS) is expected to become significant at high densities.
.

During the gravitational collapse of dark matter , due to very low
thermodynamic pressure , nearly $ 10 ^{-7} $ times that of normal
matter, there is no mechanism to halt collapse except Fermi degeneracy
pressure. 

This occurs over length scales of Compton or de Broglie
wavelengths, which are very small $ 10 ^ { -19} m $ as neutralino mass
is large. If the interparticle separation is of this order then the
mass is inside a horizon. 

The degenerate Fermi gas collapses by self
gravitation and this process is accelerated by excess boson creation as densities
increase and interparticle separation decreases. It is shown that the
Bose condensation of these Bosons accelerates the collapse in spite of
neutralino degeneracy pressure, even for
low mass dark matter lumps, thus driving the matter inside the
Schwarzschild radius. Hence a mass spectrum of dark matter black holes
is expected.

 From the studies of primordial black holes
and their mass spectrum it is expected that upto  $ 10 ^{ 15} $ kg black
holes would be unstable due to Hawking effect of quantum
evaporation. 

Larger mass black holes are classical and will survive
and grow. These will acquire discs of dark matter and perturb their
local regions in the galaxy halos. 

The larger mass black holes will
have life time of decay upto life time of galaxies. The last stage of
evaporation is rapid as most energy output occurs then. Luminosity is
$ 1/ M ^{2}$ and life time is $ M ^{1/3} $. 

This will result in normal
matter creation and gamma ray emissions. Some evidence for this in the
dark matter halos of galaxies is being found ; but better observations
are expected in a decade.

Bosons created by SUSY fermionic  Neutralino annhillation as well as
the exchange bosons in their interaction create a relativistic Bose gas, in
the high density stage. This Bose gas is mixed with the fermionic
neutralino gas. 

As the large hadron collider coming up will likely
generate data on the SUSY particles, the physical parameters will
become known and will refine the astrophysical observations.

 The statistical physics in the high density stage of dark matter
 collapse
requires an
 effective potential in Schwarzschild geometry. This can be inserted
 in place of the Newtonian gravitational potential.

 Alternately using co
 falling coordinates the $ ( 1- 2GM/rc^{2} )$ factor can be included in
 rescaled lengths along radial directions.

The conditions for the Bose condensate and the relativistic Fermi gas
to collapse into and form a  dark matter black hole are given as follows.

The non relativistic Bose condensation condition is 

$ N = \frac{V \zeta ( 3/2) } { \lambda ^{3} } $

For ultra relativistic gas $ \mu  = 0 , z=1 $ and no condensation.

 This becomes $ N = \frac{ V \zeta(3)} {\lambda^{3} } $ for
 relativistic case 

Solving for the volume of the condensate and equating it to interior volume of
the black hole we obtain a condition for formation of the black hole. 

If this number of particles resides in a sphere of radius

 $ R_{s} = \frac{ 2GM}{c^{2}}$ where $ M = M_{b} + M_{f} $ ,

 $ M_{b} = N 100 GeV $ and

$ M_ {f} = N 1000 GeV $ and

 $ V = 4/3 \pi R_{s}^{3} $ ,

 then the dark
matter black hole condition is satisfied.

The $ 10 ^{ 2} M_{s} <  M < 10 ^{8} M_{s} $

 clumps form due to
annhillation processes at the pregalactic stage and become the galaxy
core supermassive black holes. The Fermi degeneracy pressure of
neutralino gas is overcome if the condensate volume is inside the
Schwarzschild volume $ 4/3 \pi (2GM/c^{2} ) ^{3} $

\section
\qquad\qquad\qquad \qquad \qquad \qquad  CONCLUSION

There is growing research on the details of galaxy halos, formation of
lumps and black holes, as well as the nature of the neutralino gas as
dark matter. This paper attemted a simplified model of the formation
of galaxies and halos and dark matter black holes. 

The statistical
physics of normal matter ( baryons and leptons) and dark matter (
neutralinos) was developed. This provides a basic explanation of phase
separation in galaxy halos and interior of galaxies of  fermionic gases with two distinct mass species of
particles, neutralinos and baryons . 

The gravitational collapse  of low density and pressure
neutralino gas also gave rise to lumps of dark matter in the halos of
galaxies. 

Further gravitational collapse into black holes was described using Bose
condensation of exchange and annhillation bosons along with the
neutralino gas overcoming Fermi degeneracy pressure of neutralinos.

Detailed models of density functions and non spherical shapes as well
as equations of state using theories of interacting neutralinos and
partition functions are being made that will be refined with more observations.

\section
\qquad \qquad \qquad \qquad \qquad       ACKNOWLEDGEMENTS

I thank the Institute of Mathematical Sciences, Chennai for its
hospitality and excellent facilities. I thank the Director Dr Balasubramaniam and Dr Sharatchandra for supporting my
visit. Discussions with Institute members are acknowledged. I also
express my appreciation for  the students who have worked on projects guided
by me.

\section
\qquad\qquad\qquad \qquad \qquad REFERENCES

1. Lars Bergstrom, Joaakim Edsjo, Paolo Gondolo, Piero Ullio

   Clumpy neutralino dark matter. www.arxiv.org: astro-ph/9806072

2. Neven Bilic, Faustin Munyaneza, Raoul Viollier

   Stars and halos of degenerate Relativistic heavy neutrino and

   neutralino matter,astro-ph/9801262

3. R Ali Vanderveld, Ira Wasserman
  
   Spherical gravitational collapse of annhilating dark matter and the

   minimum mass of CDM black holes , astro-ph/0505497

4. hep-ph/0305047, hep-ph/0002226, hep-ph/0005171, astro-ph/0008115,

   astro-ph/0711.0466

5. astro-ph/0206036, hep-ph/0804.3084, astro-ph/0802.4348, hep-ph/0711.1240,

6. W Greiner, L.Neise, H Stocker, Thermodynamics and Statistical

   Physics. Springer 

7, Binney and Tremaine, Galaxy dynamics

8. T Padmanabhan, Theoretical Astrophysics, Cambridge university press

9. Ajay Patwardhan hep-th/0801.0304, hep-th/0705.2572,
                  
                   hep-th/0611044, hep-th/0611007, hep-th/0606080,
                  
                   hep-th/0406049,  hep-th/0310136, hep-th/0305150

\end{document}